\documentclass{article}
\usepackage{amssymb,amsmath,latexsym,amscd,amsfonts}
\usepackage{graphics}
\usepackage{epsfig}

\newtheorem{thm}{Theorem}[section]
\newtheorem{proposition}{Proposition}[section]
\newtheorem{definition}{Definition}[section]

\newtheorem{lem}{Lemma}[section]

\newenvironment{proof}{{\noindent{\bf Proof:}}}{$\hfill\Box$}

\def\NP{{\sf{NP}}}

\def\IP{{\sf{IP}}}

\def\QMA{{\sf{QMA}}}

\def\PSPACE{{\sf{PSPACE}}}

\def\hpic #1 #2 {\mbox{$\begin{array}[c]{l}
\epsfig{file=#1,height=#2} \end{array}$}}
\def\vpic #1 #2 {\mbox{$\begin{array}[c]{l}
\epsfig{file=#1,width=#2}\end{array}$}}

\begin{document}

\title{ On the Complexity of Computing Zero-Error and Holevo Capacity of Quantum Channels}
\author{Salman Beigi\thanks{salman@math.mit.edu} \hspace{1cm} Peter W. Shor\thanks{shor@math.mit.edu} \\
\normalsize\it{Department of Mathematics, Massachusetts Institute
of Technology, Cambridge, MA } }


\maketitle{}

\begin{abstract}

One of the main problems in quantum complexity theory is that our
understanding of the theory of $\QMA$-completeness is not as rich
as its classical analogue, the $\NP$-completeness. In this paper
we consider the clique problem in graphs, which is $\NP$-complete,
and try to find its quantum analogue. We show that, quantum
clique problem can be defined as follows; Given a quantum
channel, decide whether there are $k$ states that are
distinguishable, with no error, after passing through channel.
This definition comes from reconsidering the clique problem in
terms of the zero-error capacity of graphs, and then redefining
it in quantum information theory. We prove that, quantum clique
problem is $\QMA$-complete.

In the second part of paper, we consider the same problem for the
Holevo capacity. We prove that computing the Holevo capacity as
well as the minimum entropy of a quantum channel is
$\NP$-complete. Also, we show these results hold even if the set
of quantum channels is restricted to entanglement breaking ones.

\end{abstract}

\section{Introduction}

One of the basic results of complexity theory is Cook-Levin
theorem. That is, SAT, the problem of whether a Boolean formula
has a satisfying assignment or not, is $\NP$-complete. In fact,
the Cook-Levin theorem was the beginning of the theory of
$\NP$-completeness. After SAT, a series of natural problems in
graph theory and combinatorics were shown to be $\NP$-complete as
well, see \cite{sipser}. Hamiltonian cycle, clique problem, graph
coloring, subset sum and vertex cover are examples of such
problems. The rich theory of $\NP$-completeness has been used in
other parts of complexity theory as well. For instance, the basic
ideas of such important results as $\IP=\PSPACE$ \cite{shamir},
and the PCP theorem \cite{arora}, are from this theory.

In quantum complexity theory, Kitaev's result \cite{kitaev} is
considered as the quantum analogue of Cook-Levin theorem. The
complexity class $\QMA$ was defined by Watrous as the quantum
analogue of $\NP$. $\QMA$ is the class of problems that can be
solved by a quantum polynomial time algorithm given a quantum
witness. Then Kitaev defined a natural problem in physics, called
the {\it local Hamiltonian problem} and showed it is
$\QMA$-complete, \cite{kitaev}. Indeed, the local Hamiltonian
problem is considered as the quantum version of SAT problem, and
also Kitaev's result as the quantum analogue of Cook-Levin
theorem. Thus, same as in classical case, one would guess that
this is the beginning of a rich theory for $\QMA$-complete
problems.

After Kitaev's theorem few problems have been shown to be
$\QMA$-complete. For instance, circuit identity testing
\cite{janzing} and local consistency problem \cite{liu} are such
problems. But this class is not as rich as the class of
$\NP$-complete problems.

One of the most natural ways of thinking of $\QMA$-complete
problems is to somehow define the quantum version of problems
that are known to be $\NP$-complete. An example of such a problem
is the clique problem in graphs.

\subsection{Quantum-clique problem}\label{subsec1}

In a graph $G$ a {\it clique} is a subset of vertices every two of
which are adjacent, and the size of a clique is the number of its
vertices. The clique problem is that given a graph $G$ and an
integer number $k$, decide whether $G$ contains a clique of size
$k$ or not. It is well-known that clique problem is
$\NP$-complete, \cite{sipser}.

We can think of this problem in $G^c$, the complement of graph
$G$. In the complement of $G$ a clique is changed to an {\it
independent set}. An independent set in a graph is a subset of
vertices no two of which are adjacent, and the maximum size of an
independent set is called the {\it independence number} of $G$,
denoted $\alpha(G)$. So the clique problem in the complement
graph reduces to decide whether $\alpha(G)\geq k$, and then it is
$\NP$-complete.

This reduction is important because the problem of computing
$\alpha(G)$ is related to the problem of computing the {\it
zero-error capacity} of a channel. Zero-error capacity of a
classical (discrete memoryless) channel is the maximum rate of
information that one can send through channel without error. This
concept is first introduced by Shannon in his famous paper
\cite{shannon}, and got much attention after that. For a survey
on this topic see \cite{korner}. Also some interesting results on
graph capacity are presented in \cite{alon,haemers,lov}.

Indeed, any classical channel corresponds to a graph $G$, and the
zero-error capacity of the channel can be computed in terms of the
independence number of $G$ and its powers. We will get to this
correspondence latter on, but here we mention that the best way
of coding messages in words of length one is to find an
independent subset of size $\alpha(G)$. Therefore it is
$\NP$-complete to find such a codeword.

This way of thinking of clique problem is useful because it is in
terms of channels and their zero-error capacity, the concepts
that are already known in quantum information theory. The
definition of a quantum channel is well-known in the theory. Also
extending the definition of zero-error capacity to quantum
channels is straightforward, \cite{jor}. That is, what is the
maximum rate of classical information that can be sent through a
quantum channel with zero-error and without using entanglement.
Thus, the same question as in the classical case arises; How can
we compute the zero-error capacity of a quantum channel?

As in classical case let us first try to code the messages in
words of length one (not in product states). Then the question is
that, are there $k$ states that after passing through channel can
be recognized with no error? This is exactly the quantum analogue
of the clique problem. So we call it the {\it quantum clique
problem}. In this paper we prove that quantum clique is
$\QMA$-complete.

\subsection{Computing the channel capacity}

In the previous section we said that the problem of estimating
the zero-error capacity of a classical channel (graph) is
$\NP$-complete, and its quantum analogue is $\QMA$-complete. Both
of these problems are about the channel capacities in the
zero-error case. But, how hard is to compute the usual capacity
of a channel?

In the classical case, there is an algorithm, called {\it
Arimoto-Blahut algorithm}, that computes the capacity of a
classical channel efficiently, \cite{arimoto,blahut}. In the
quantum case, the same idea as in Arimoto-Blahut algorithm leads
us to an algorithm for computing the Holevo capacity, see
\cite{nagaoka,osawa}. The idea is to use Klein's inequality,
\cite{chuang}, iteratively and find a sequence that converges to a
local maximum of the expression in the Holevo capacity. Although
finding this sequence can be done in polynomial time, it gives a
local maximum not a global one, and unlike the classical case a
local maximum is not a global one. So, the quantum version of
Arimoto-Blahut algorithm fails. Here, we show that finding a bound
for the Holevo capacity of a quantum channel is $\NP$-complete.

To prove this result we first show that computing the minimum
entropy of a quantum channel is $\NP$-hard and then convert the
problem of computing the Holevo capacity to this one.

\section{Preliminaries}

\subsection{$\QMA$} \label{subsec:qma}

\begin{definition}\label{qma} A language $L$ is said to be in $\QMA=\QMA(2/3, 1/3)$ if there
exists a quantum polynomial time verifier $V$ such that

\begin{itemize}
\item Completeness: $\forall x\in L, \exists\, \vert \xi\rangle$,
$Pr(V(\vert x\rangle \vert \xi\rangle)\ accepts)\geq 2/3$.

\item Soundness: $\forall x\notin L, \forall \vert \xi\rangle$, $Pr(V(\vert x\rangle
\vert \xi\rangle)\ accepts)\leq 1/3$.
\end{itemize}

\end{definition}

An easy amplification argument implies that the completeness and
soundness bounds, $2/3$ and $1/3$, are not crucial and can be
replaced by any functions $a(n)$ and $b(n)$ provided that, they
are different from $0$ and $1$ by an inverse exponential
function, and also there is an inverse polynomial gap between
them. In other words, if $a(n)$ and $b(n)$ are two functions such
that $0< b(n)< a(n)<1$, and for some constant $c$,
$$a(n)<1-e^{-n^c},\ \ \ b(n)> e^{-n^c},$$
and $$a(n)-b(n)<n^{-c},$$ then $\QMA(a, b)=\QMA(2/3,1/3)=\QMA$,
see \cite{survey}.

\begin{definition} $k$-Local Hamiltonian problem $( H_1,\dots H_s, a ,b)$

\begin{itemize}
\item {\bf Input:} An integer $n$, real numbers $a, b$ such that $ b-a > n^{-c}$, and
polynomially many Hermitian non-negative semidefinite matrices $H_1,\dots H_s$ with bounded norm,
$\parallel H_i\parallel \leq 1$, such that each of them acts just on $k$ of $n$ qubits.

\item {\bf Promise:} The smallest eigenvalue of $H_1+\dots H_s$ is either less than $a$
or greater than $b$.

\item {\bf Output:} Decide which one is the case.
\end{itemize}

\end{definition}

Kitaev proved that local Hamiltonian problem for $k=5$ is
$\QMA$-complete \cite{kitaev}, but latter this result was improved
in \cite{kempe2} and \cite{kempe1}.

\begin{thm} \label{thm:hamiltonian} $2$-local Hamiltonian problem is
$\QMA$-complete.

\end{thm}

\subsection{$\QMA_1$} \label{subsec:qma1}

We said that $\QMA(a, b)$ is equal to $\QMA(2/3, 1/3)$ if $a\neq 1
,\, b\neq 0$, and there is a polynomial gap between them. But one
may ask about the case where $a=1$, or $b=0$, i. e. perfect
completeness or soundness. For instance $\QMA(1,1/3)$, is set of
languages that have a protocol as in definition \ref{qma} with
completeness bound $1$ and soundness bound $1/3$. Same as before,
an amplification argument shows that if $b(n)$ is such that
$e^{-n^c}<b(n)<1-n^{-c}$, for some constant $c$, then $\QMA(1,
b)=\QMA(1, 1/3)$. Therefore, we get to an enough robust
complexity class, denoted $\QMA_1$, where the subscript $1$ shows
the perfect completeness.

This complexity class was first introduced by Bravyi
\cite{bravyi}. He also defined the {\it quantum $k$-SAT problem}
which is a special case of local Hamiltonian problem. To state
this problem precisely we need to fix some notions. In the Hilbert
space of $n$-qubits, a projection is a Hermitian operator $\Pi$
such that $\Pi^2=\Pi$, i.e. eigenvalues of $\Pi$ are $0$ and $1$.
Also, we say $\Pi$ is a $k$-projection if it acts just on $k$
qubits.

\begin{definition}\label{def:qsat} Quantum $k$-SAT problem $(\Pi_1,\dots \Pi_s, \epsilon)$

\begin{itemize}
\item {\bf Input}: An integer $n$, a real number $\epsilon>
n^{-c}$, and polynomially many $k$-projections $\Pi_1, \dots
\Pi_s$.

\item {\bf Promise}: Either there exists an $n$-qubit state $\vert \psi\rangle$
such that $\Pi_i\vert \psi\rangle=0$ for all $i$, or $\sum_i\,
\langle \psi\vert \Pi_i\vert\psi\rangle \geq \epsilon$ for all
$\vert \psi\rangle $.
\item {\bf Output}: Decide which one is the case.
\end{itemize}

\end{definition}

Given a witness $\vert \psi\rangle$, in polynomial time we can
decide whether $\Pi_i\vert \psi\rangle=0$ or not, and then
$k$-SAT is in $\QMA_1$. But we should be careful because we need
a protocol that has no error if $\vert \psi \rangle$ is the right
witness (the common eigenvector of $\Pi_i$'s). It means that the
verifier should be able to check $\Pi_i\vert \psi\rangle=0$ with
no error, or equivalently, the verifier should be able to
implement the projection $\Pi_i$ exactly.

This extra condition on the verifier for $\QMA_1$ protocols
arises naturally not only for quantum $k$-SAT, but also for any
other problem in this class. If we can implement the gates just
with an approximation, then our algorithm contains some error
anyway. This problem can be resolve by emphasizing that the
verifier can implement all quantum gates up to three-qubit gates,
exactly. So in this paper, by a quantum verifier for $\QMA_1$
protocols we mean the one that has all three-qubit quantum gates
in hand. Bravyi pointed out this assumption in \cite{bravyi} and
proved the following lemma.

\begin{lem} \label{lem:3gate} Let $U$ be a unitary operator acting on k qubits.
Then $U$ can be exactly represented by a quantum circuit of size
$poly(k)2^{2k}$ with three-qubit gates.
\end{lem}

Also he showed the following theorem, that has almost the same
proof as theorem \ref{thm:hamiltonian}.

\begin{thm} \label{sat} Quantum $4$-SAT is
$\QMA_1$-complete.

\end{thm}

Both theorems \ref{thm:hamiltonian} and \ref{sat} are important
in the theory of $\QMA$-complete problems. Because if we show some
problem is in $\QMA$, ($\QMA_1$) and also find a reduction from
local Hamiltonian (quantum SAT) to it then we conclude that it is
$\QMA$-complete ($\QMA_1$-complete).

\subsection{Zero-error channel capacity}

A classical discrete memoryless channel consists of an input set
$X$, an output set $Y$, and probability distributions $p(y\vert
x)$ for every $x\in X$ and $y\in Y$, meaning that if we send $x$
through channel we get $y$ as output with probability $p(y\vert
x)$. Since, we want to define the zero-error capacity of this
channel, the exact value of $p(y\vert x)$ is not important for
us, but whether it is zero or not. Therefore, to get a clearer
representation, we correspond to the channel a graph $G$ on the
vertex set $X$ in which two vertices $x,x'\in X$ are adjacent if
there is $y\in Y$ such that $p(y\vert x), p(y\vert x')$ are both
non-zero. It means that, $x, x'$ are adjacent in $G$ if they can
be confused after passing through channel. Hence, some messages
$x_1,\dots x_k$ can be sent through channel with no error iff
there is no edges between them, i. e. $\{x_1, \dots x_k\}$ is an
independent set.

\begin{definition} In a graph $G$, a subset of vertices no two of which are adjacent is called
an independent set. Also $\alpha(G)$ denotes the maximum size of
an independent set in $G$.
\end{definition}

By the above discussion, if we want to code our messages in words
of length one (one use of channel) then the best way is to code
them in an independent set of maximum size. In this case we get
to the rate $\alpha(G)$. But we may use words of length two. In this case we get to another
graph, denoted  $G\otimes G$.

\begin{definition} Assume $G$ and $H$ are two graphs on vertex sets
$V$ and $U$, respectively. Then their tensor product $G\otimes H$
is a graph on the vertex set $V\times U$ such that $(v_1, u_1)$
and $(v_2, u_2)$ are adjacent if $v_1$, $v_2$ are either equal or
adjacent in $G$, and also $u_1, u_2$ are either equal or adjacent
in $H$.
\end{definition}

It is not hard to see that the graph corresponding to words of
length two is $G\otimes G$. Thus, the best way to code the
messages in words of length two is to use an independent set in
$G\otimes G$ of size $\alpha(G\otimes G)$. So we get to the rate
$\alpha(G\otimes G)^{1/2}$ (square root is for normalization).
Repeating this argument for higher products, we get to the
following definition due to Shannon \cite{shannon}.

\begin{definition} $\Theta(G)$, the capacity of the graph $G$, is equal
to $$\Theta(G)=\lim_{n\rightarrow\infty}\, \alpha(G^{\otimes
n})^{\frac{1}{n}}.$$

\end{definition}

These definitions all can be generalized for quantum channels,
\cite{jor,med}. The zero-error capacity of a quantum channel is
the maximum rate of classical information that one can send
through a quantum channel without using entanglement. To get a
closed form expression for this quantity, suppose $\Phi$ is a
quantum channel, and we code $k$ messages in quantum states
$\rho_1,\dots \rho_k$. If we want to decode the outputs of
channel with no error, we should be able to identify states
$\Phi(\rho_1),\dots \Phi(\rho_k)$ without error. It is well-known
that some quantum states can be recognized with no error iff they
have orthogonal supports.

\begin{definition} For a quantum channel $\Phi$, $\alpha(\Phi)$ is
the maximum number of states $\rho_1, \dots \rho_k$ such that
$\Phi(\rho_1), \dots \Phi(\rho_k)$ have orthogonal supports.

Also, $\alpha(\Phi^{\otimes n})$ is the maximum number of product
states $\rho_{i1}\otimes \dots \otimes\rho_{in}$, $i=1,\dots k$,
such that all states $\Phi^{\otimes n}(\rho_{i1}\otimes \dots
\otimes\rho_{in})$, $i=1,\dots k$, have orthogonal supports.

\end{definition}

To make it clear we should highlight two points. First of all, if
$\Phi(\rho)$ and $\Phi(\rho')$ have orthogonal supports then
$\rho$ and $\rho'$ also have orthogonal supports. Therefore
$\alpha(\Phi)$ is at most the dimension of input states, and is
finite. Second, we emphasize that the input states of
$\Phi^{\otimes n}$ should be product states (because we do not
want to use entanglement) and by abuse of notation we denote the
maximum number of such product states by $\alpha(\Phi^{\otimes
n})$.

By the above definition it is clear what the zero-error capacity
of a quantum channel should be.

\begin{definition} $\Theta(\Phi)$, the zero-error capacity of the
quantum channel $\Phi$, is
$$\Theta(\Phi)=\lim_{n\rightarrow\infty}\, \alpha(\Phi^{\otimes
n})^{\frac{1}{n}}.$$

\end{definition}

\subsection{How to compute $\alpha(\Phi)$}\label{subsec:computealpha}

We do not repeat all the known properties of $\alpha(\Phi)$ and
refer the reader to \cite{jor}. Here, we need just two basic
properties. Suppose $\alpha(\Phi)=n$, and $\rho_1,\dots \rho_n$
are $n$ states such that the supports of $\Phi(\rho_1),\dots
\Phi(\rho_n)$ are orthogonal. For $i=1, \dots n$, let $\vert
\psi_i\rangle$ be a pure state in the support of $\rho_i$. Then,
since the support of $\Phi(\vert \psi_i\rangle)$ is a subspace of
the support of $\Phi(\rho_i)$, the states $\Phi(\vert
\psi_1\rangle),\dots \Phi(\vert \psi_n\rangle) $ have orthogonal
supports as well. It means that, to compute $\alpha(\Phi)$ it
suffices to restrict ourselves to pure states.

Now assume that the operator sum representation of $\Phi$ is

\begin{equation}\label{osr} \Phi(\rho)=\sum^r_{k=1} E_k\rho E^\dagger_k,
\end{equation}
where $\sum^r_{k=1} E^\dagger_kE_k=I$. Then the support of
$\Phi(\vert \psi_i\rangle)$ is spanned by vectors $E_1\vert
\psi_i\rangle,\dots E_r\vert \psi_i\rangle$. Therefore $\Phi(\vert
\psi_1\rangle),\dots \Phi(\vert \psi_n\rangle) $ have orthogonal
supports iff these vectors, for different indices $i$ and $j$, are
orthogonal. Summarizing these two statements, we get to the
following proposition.

\begin{proposition} For a quantum channel $\Phi$ with operator sum
representation (\ref{osr}), we have $\alpha(\Phi)\geq n$ if and only if
there exist pure states $\vert \psi_1\rangle, \dots \vert
\psi_n\rangle$ such that $\langle \psi_i \vert E^\dagger_k
E_l\vert \psi_j\rangle=0$, for every $k,l$ and  $i,j$, where
$i\neq j$.

\end{proposition}

\subsection{Quantum clique problem}

We know that deciding whether a given graph has a clique of size
$k$ is $\NP$-complete. Considering this problem in the complement
graph we find that, deciding whether $\alpha(G)\geq k$ is
$\NP$-complete. In our notion, it means that having a classical
channel, deciding whether by coding messages in words of length
one, we can get to the rate $k$ for transmitting information with
zero-error, is $\NP$-complete. Since we have all these notions
for the quantum case we can define the {\it quantum clique
problem}.

Basically, the quantum version of clique problem is also to
decide whether $\alpha(\Phi)\geq k$, for a given quantum channel
$\Phi$. It is equivalent to decide whether there exist quantum
states $\rho^1,\dots \rho^k$ such that $\Phi(\rho^1),\dots
\Phi(\rho^k)$ have orthogonal supports or not. Note that, for any
two states $\sigma^1, \sigma^2$, we have $tr(\sigma^1\sigma^2)\geq
0$ and equality holds if and only if $\sigma^1, \sigma^2$ have
orthogonal supports.

Let $\sigma^{1,2}=\sigma^1\otimes \sigma^2$ then
$tr(\sigma^1\sigma^2)=tr(S\,\sigma^{1,2})$, where $S$ is the swap
gate ($S\vert \psi\rangle\vert\phi\rangle=\vert \phi\rangle\vert
\psi\rangle$). Therefore by applying the swap gate we can estimate
$tr(\sigma^1\sigma^2)$. But notice that if $\sigma^{1,2}$ is not
separable then this equality does not hold and the orthogonality
of $\sigma^1$ and $\sigma^2$ is not implied by
$tr(S\,\sigma^{1,2})=0$. To resolve this problem we can restrict
ourselves to {\it entanglement breaking channels} to ensure that
the output states of the channel are not entangled.

A quantum channel $\Phi$ is called entanglement breaking if there
are POVM $\{M_i\}$ and states $\sigma_i$ such that
$$\Phi(\rho)=\sum_i tr(M_i\rho)\sigma_i .$$ In this case,
$\Phi^{\otimes 2}(\rho^{1,2})$ is always separable, $tr(S\,
\Phi^{\otimes 2}(\rho^{1,2}))\geq 0$ and equality implies
$\Phi(\rho^1)$ and $\Phi(\rho^2)$ are orthogonal.

\begin{definition} Quantum clique problem $(\Phi, k, a, b)$

\begin{itemize}

\item {\bf Input} Integer numbers  $n$ and $k$, non-negative real numbers $a, b$ with an inverse
polynomial gap $b-a > n^{-c}$, and an entanglement breaking
channel $\Phi$ that acts on $n$-qubit states.

\item {\bf Promise} Either there exists $\rho^1\otimes \dots \otimes\rho^k$ such that
$\sum_{i,j} tr(S\,\Phi(\rho^i)\otimes\Phi(\rho^j)) \leq a $ or for
any state $\rho^{1,2\dots k}$ we have $\sum_{i,j}
tr(S\,\Phi^{\otimes 2}(\rho^{i,j})) \geq b$.

\item {\bf Output} Decide which one is the case.

\end{itemize}

\end{definition}

Notice that, if we let $a=0$ we get to the exact orthogonality
assumption that is a special case of quantum clique, and in
general is a simpler problem. Indeed, we show that quantum clique
problem is $\QMA$-complete, and in the special case where $a=0$
and $\Phi$ is restricted to quantum-classical channels, it is
$\QMA_1$-complete.

\subsection{Holevo capacity}

The Holevo capacity of a quantum channel is the maximum rate of
classical information that can be sent through a quantum channel
without using entanglement, \cite{chuang}. Assume that $\Phi$ is a
quantum channel. Then $\chi(\Phi)$, the Holevo capacity of
$\Phi$, is equal to

\begin{equation}\label{holevo} \chi(\Phi) =\max_{p_i, \rho_i}\ H(\sum_i p_i\Phi(\rho_i))
-\sum_i p_i H(\Phi(\rho_i)),
\end{equation}
where $H(\rho)=-tr (\rho\log \rho)$ denotes the {\it von Neumann}
entropy, and the maximum is taken over probability distributions
$\{p_i\}$ and quantum states $\{\rho_i\}$. Using the convexity of
von Neumann entropy, we can assume that states $\rho_i$ are pure.
Also, if $\Phi$ acts on an $n$-dimensional Hilbert space then we
may assume that number of $\rho_i$'s is at most $n^2$,
\cite{chuang}. However, these are not enough information on what
the maximum point is, and how we can compute $\chi(\Phi)$.

There is an algorithm called the Arimoto-Blahut algorithm that
given a classical discrete memoryless channel computes its
capacity, see \cite{arimoto,blahut}. Indeed, computing the
capacity of a classical channel involves maximization of some
mutual information. In the Arimoto-Blahut algorithm this
maximization problem is converted to an alternating maximization
one, that tends to the channel capacity and is more tractable.
Using the same idea, Nagaoka in \cite{nagaoka} proposed the same
algorithm to compute the Holevo capacity . But, the point is that
in (\ref{holevo}) there can be a local maximum which is not a
global one. So that, in the quantum Arimoto-Blahut algorithm the
alternate maximum value may tend to a local maximum, not to
$\chi(\Phi)$.

In this paper, we prove that computing $\chi(\Phi)$ is
$\NP$-complete. In fact, a more strong theorem holds: computing
the Holevo capacity of entanglement breaking channels is
$\NP$-complete.

\subsection{Minimum entropy of a quantum channel}

Minimum entropy of a quantum channel is equal to the minimum
entropy of its output states,

\begin{equation}\label{eq:min}\min_{\rho}
H(\Phi(\rho)).
\end{equation}

Again, using the convexity of von Neumann entropy, the minimum is
achieved on pure states. The minimum entropy is an important
invariant of quantum channels. Indeed, it is proved that the
famous additivity conjecture of Holevo capacity is equivalent to
the additivity of minimum entropy, \cite{shor}. This result is
important for us because it somehow expresses the minimum entropy
in terms of Holevo capacity, and using this idea we convert the
problem of computing the minimum entropy to the problem of
computing Holevo capacity. Indeed, to prove that computing Holevo
capacity is $\NP$-complete, we first state the $\NP$-completeness
of computing minimum entropy.

\subsection{SWAP test}\label{subsec:swap}

SWAP test is a well-know protocol for deciding whether two given
quantum states are the same or not. The protocol is as follows.
Given two states $\vert \psi_1\rangle$ and $\vert \psi_2\rangle$
and an ancilla qubit $\vert 0\rangle$, first apply the Hadamard
gate on the ancilla, then the controlled-swap gate on two
registers, and again, Hadamard on the ancilla. At the end, measure
the ancilla qubit in the computational basis. It is easy to see
that this protocol computes the channel

\begin{equation}\label{eq:swaptest}
\Phi_{swap}(\vert \psi_1\rangle \vert \psi_2\rangle)=
\frac{1}{2}(1+\vert \langle \psi_1\vert \psi_2\rangle\vert^2)
\vert  0  \rangle\langle  0  \vert + \frac{1}{2}(1-\vert \langle
\psi_1\vert \psi_2\rangle\vert^2) \vert  1 \rangle\langle 1 \vert
.
\end{equation}
In fact, in the measurement we get $\vert 0\rangle$ with
probability $\frac{1}{2}(1+\vert \langle \psi_1\vert
\psi_2\rangle\vert^2)$ and $\vert 1\rangle$ with probability
$\frac{1}{2}(1-\vert \langle \psi_1\vert \psi_2\rangle\vert^2)$.
Therefore, if we correspond the output $\vert 0\rangle$ to $+1$
and output $\vert 1\rangle$ to $-1$ then the expected value of
this number is equal to $\vert \langle \psi_1\vert
\psi_2\rangle\vert^2$. In general, when the input state is
$\sigma^{12}$ we can compute $tr(S\,\sigma^{12})$, where $S$ is
the swap gate ($S\vert \psi_1\rangle\vert \psi_2\rangle= \vert
\psi_2\rangle\vert \psi_1\rangle$).

\section{Complexity of quantum clique problem}

\subsection{Quantum clique is $\QMA$-complete}

Here is the main theorem of this section.

\begin{thm} \label{thm:qclique} The quantum clique problem $(\Phi, k, a, b)$ where $\Phi$
is an entanglement breaking channel on $n$-qubit states and has
the operator sum representation
\begin{equation} \label{eq:osr2}
\Phi(\rho)=\sum_{i=1}^{r} E_i\rho E_i^\dagger,
\end{equation}
where $\sum_i E_{i}^\dagger E_i=I$ and $r=poly(n)$, is $\QMA$-complete.
\end{thm}

\begin{proof} First we show that $(\Phi, k, a, b)$ is in $\QMA$. Note that, $\Phi$ can be
written as $\Phi(\rho)=tr_2\big( U\rho\otimes \vert
1\rangle\langle 1\vert U^\dagger\big)$, where $U$ is a unitary
operator and
\begin{equation} \label{eq:u}
U\vert \psi\rangle\vert 1\rangle=\sum_{i=1}^{r}\ E_i\,\vert \psi
\rangle\, \vert i\rangle.
\end{equation}

Since $r=poly(n)$, a polynomial time verifier can implement $U$
and then $\Phi$, with arbitrary small error. Therefore, given
witness $\rho^{1,\dots k}$, verifier can randomly choose $i,j$,
$1\leq i,j\leq k$, compute $\Phi^{\otimes 2}(\rho^{i,j})$, and
then apply the SWAP test. As we said in section
\ref{subsec:swap}, the expected value of the outcome of SWAP test
for fixed $i,j$, is $tr\big(S\, \Phi^{\otimes 2
}(\rho^{i,j})\big)$, and for random choices of $i,j$ is equal to
$$\frac{1}{{k \choose 2}}\sum_{i,j}\ tr\big(S\, \Phi^{\otimes 2
}(\rho^{i,j})\big),$$ which is either less than
$\frac{2}{k(k-1)}a$ or greater than $\frac{2}{k(k-1)}b$. Hence,
there is an inverse polynomial gap between them and the verifier
can recognize them in polynomial time. Thus, quantum clique
problem is in $\QMA$.

To prove the hardness, we establish a polynomial time reduction
from local Hamiltonian problem to quantum clique. Let $(H_1,\dots
H_s, a, b)$ be an instance of local Hamiltonian problem. Since
$\parallel H_i \parallel \leq 1$, then $\frac{1}{s}H \leq I$,
where $H=\sum_i H_i$. Thus, $M=I-\frac{1}{s}H$ is a positive
operator and we can define the following quantum channel
$$\Phi(\rho)=\frac{1}{s} tr(H\otimes I\,\rho)\vert 00\rangle\langle 00\vert
            +tr\big(M\otimes \vert 0\rangle\langle 0\vert\,\rho\big) \vert 11\rangle\langle 11\vert
            +tr\big(M\otimes \vert 1\rangle\langle 1\vert\,\rho\big) \vert 10\rangle\langle 10\vert.   $$
Note that, $s=poly(n)$ and then $\Phi$ is of the form of
(\ref{eq:osr2}), ($r=poly(n)$). So, we can consider $(\Phi,
k=2,\frac{1}{s^2}a^2 ,\frac{1}{s^2}b^2 )$ as an instance of
quantum clique. We prove that $(H_1,\dots H_s, a, b)$ is a "yes"
instance of local Hamiltonian if and only if $(\Phi,
k=2,\frac{1}{s^2}a^2 ,\frac{1}{s^2}b^2 )$ is a "yes" instance of
quantum clique.

Suppose $(H_1,\dots H_s, a, b)$ is a "no" instance. Then for any
state $\sigma$, $tr(H\sigma)\geq b$, and then, for any state
$\rho^{1,2}$ we have

$$tr\big(S\, \Phi^{\otimes 2
}(\rho^{1,2})\big)\geq \frac{1}{s^2} tr(H\otimes
I\,\rho^1)tr(H\otimes I\,\rho^2) \geq \frac{1}{s^2}b^2.$$ So,
$(\Phi, k=2,\frac{1}{s^2}a^2 ,\frac{1}{s^2}b^2 )$ is also a "no"
instance. Now, assume that there is $\vert \psi\rangle$ such that
$\langle \psi\vert H\vert \psi\rangle\leq a$. Let $\rho^1=\vert
\psi\rangle \langle \psi\vert \otimes \vert 0 \rangle\langle
0\vert $, and $\rho^2= \vert \psi\rangle \langle \psi\vert \otimes
\vert 1 \rangle\langle 1\vert$. We have

\vspace{.4cm}

$tr\big(S\, \Phi(\rho^1)\otimes\Phi(\rho^2)\big)$  $$= tr\Big(
(\frac{1}{s}\langle \psi\vert H\vert \psi \rangle \vert 00\rangle
\langle 00\vert +  \langle \psi\vert M\vert \psi \rangle \vert
11\rangle \langle 11\vert )    (\frac{1}{s}\langle \psi\vert
H\vert \psi \rangle \vert 00\rangle \langle 00\vert +  \langle
\psi\vert M\vert \psi \rangle \vert 10\rangle \langle 10\vert
)        \Big)$$
$$=\frac{1}{s^2}\langle \psi\vert H\vert \psi\rangle ^2\leq \frac{1}{s^2}a^2.$$
Therefore, $(\Phi, k=2,\frac{1}{s^2}a^2 ,\frac{1}{s^2}b^2 )$ is
also a "yes" instance.

\end{proof}

\subsection{Channels that can be implemented exactly}

Theorem \ref{thm:qclique} says that quantum clique problem
$(\Phi, k, a, b)$ is $\QMA$-complete. In this $\QMA$ protocol
since, in general, $a$ is a positive number, we are allowed to
have some probability of error. But one may consider the case
$a=0$ and try to find a protocol with no error. Recall that, if
$a=0$ then $(\Phi, k, a=0, b)$ exactly says that whether
$\alpha(\Phi)\geq k$ or not. Here, we show that this problem is
$\QMA_1$-complete.

The first step toward proving such a result is to show that if
$a=0$ then quantum clique is in $\QMA_1$ . But, $\QMA_1$ consists
of problems that have a quantum Merlin-Arthur prorocol with {\it
one sided error}, and in fact perfect completeness. So, we should
be able to check the orthogonality of two quantum states without
error. But, in general, the SWAP-test, that we applied in theorem
\ref{thm:qclique}, contains some non-zero probability of error.

The idea to resolve this problem is to restrict ourselves to the
special case of quantum-classical channels (q-c channels). A
channel $\Phi$ is called a q-c channel if it can be written in the
form
\begin{equation}\label{eq:q-c}
\Phi(\rho)=\sum_{i=1}^r tr(M_i\rho)\vert i\rangle \langle i\vert,
\end{equation}
where $\{M_1,\dots M_r\}$ is a POVM and $\vert 1\rangle,\dots
\vert r\rangle$ are orthogonal states. Checking orthogonality of
two outcome states of these channels is easy. Given two such
states $\Phi(\rho)$, $\Phi(\rho')$, we can measure them in the
basis $\vert 1\rangle,\dots \vert r\rangle$. If the outcome of
the measurements were the same then their supports are not
orthogonal.

Another restriction that we should keep in mind is that, the
verifier should be able to compute $\Phi(\rho)$, exactly. If the
verifier could just implement $\Phi$ with some approximation,
then all the computation contains some probability of error. In
fact, this is the same kind of restriction that we mentioned for
the quantum $4$-SAT problem. So, we should restrict the set of
channels to the quantum channels that can be implemented exactly
by a polynomial time quantum verifier. In section
\ref{subsec:qma1} we pointed out that by a quantum verifier for
$\QMA_1$ protocols we mean the one that can implement all
$3$-qubit quantum gates, exactly. But, it does not mean that we
can implement any channel with no error. So, we should clarify
that, in this case, by a quantum channel we mean the one the can
be implemented exactly by a quantum verifier. We do not need to
classify all of these channels. We just need to show that this
class of channels is enough rich.

\begin{lem} \label{lem:exact}
\hspace{1cm}

\begin{itemize}
\item[\rm{(i)}] Any quantum channel that acts just on a constant number of qubits and
has an operator sum representation of the form (\ref{eq:osr2})
where $r$ is a constant can be implemented with no error.

\item[\rm{(ii)}] For polynomially many channels $\Phi_1, \dots \Phi_s$ that can be implemented
exactly, $\frac{1}{s}\sum_i \Phi_i$ can be implemented with no
error.

\item[\rm{(iii)}] If $\Pi$ is a $k$-projection where $k$ is a constant, then the following channel can be
implemented exactly
$$ \Phi(\rho)= tr\big(\Pi\otimes I\, \rho\big)\vert 00\rangle\langle 00\vert
              +tr\big((I-\Pi)\otimes \vert 0\rangle\langle 0\vert\,\rho\big) \vert 11\rangle\langle 11\vert
              +tr\big((I-\Pi)\otimes \vert 1\rangle\langle 1\vert\,\rho\big) \vert 10\rangle\langle 10\vert.$$

\item[\rm{(iv)}] Suppose $\Pi_1, \dots \Pi_s$ are polynomially many $k$-projections, where $k$ is a constant. Let
$\Pi=\sum_i \Pi_i$, and $M=I-\frac{1}{s}\Pi$. Then the following channel can be implemented exactly.
$$\Phi(\rho)= \frac{1}{s}tr\big(\Pi\otimes I\, \rho\big)\vert 00\rangle\langle 00\vert
              +tr\big(M\otimes \vert 0\rangle\langle 0\vert\,\rho\big) \vert 11\rangle\langle 11\vert
              +tr\big(M\otimes \vert 1\rangle\langle 1\vert\,\rho\big) \vert 10\rangle\langle 10\vert.$$

\end{itemize}

\end{lem}

\begin{proof} {\rm (i)} The idea is same as what we did in the proof of theorem
\ref{thm:qclique}. In fact, such a channel can be written of the
form $\Phi(\rho)=tr_2(U\rho U^\dagger)$, where $U$ is given by
equation (\ref{eq:u}). In this special case, $U$ acts just on
constant number of qubits, and then by lemma \ref{lem:3gate} it
can be implemented efficiently and with no error.

\noindent{\rm (ii)} Pick a random $i$, $1\leq i\leq s$, and apply
$\Phi_i$.

\noindent{\rm (iii)},{\rm (iv)} are easy consequences of {\rm (i)}
and {\rm (ii)}.

\end{proof}

\subsection{$a=0$}

\begin{thm} \label{thm:qclique1} Quantum clique problem $(\Phi, k, a=0, b)$, where
$\Phi$ is a q-c channel that can be implemented exactly by a
polynomial time verifier is $\QMA_1$-complete.

\end{thm}

\begin{proof} First we show that this problem is in $\QMA_1$. Given a channel $\Phi$ of the form
(\ref{eq:q-c}), if $\alpha(\Phi)\geq k$, then there are states
$\rho^1, \dots ,\rho^k$ such that $\Phi(\rho^1),\dots
\Phi(\rho^k)$ have orthogonal supports. Hence, $\rho^1\otimes
\dots \otimes \rho^k$ is a witness, and verifier can randomly
choose two indices $i,j$, $1\leq i,j\leq k$, apply $\Phi$ on
$\rho^i$ and $\rho^j$, and then check whether $\Phi(\rho^i)$ and
$\Phi(\rho^j)$ are orthogonal or not. Since $\Phi$ is a q-c
channel, $\Phi(\rho^i)$ and $\Phi(\rho^j)$ are orthogonal iff the
outcome of their measurement in the basis $\vert 1\rangle, \dots
\vert r\rangle$, never be the same. Note that, conditioned on
$i,j$, the probability of a collision in the measurement is equal
to $tr\big(\Phi(\rho^i)\Phi(\rho^j)\big)=tr\big(S\, \Phi^{\otimes
2 }(\rho^{i,j})\big)$, and in general it is
\begin{equation} \label{eq:exp}
\frac{1}{{2\choose k}}\sum_{i,j} tr\big(S\, \Phi^{\otimes 2
}(\rho^{i,j})\big).
\end{equation}
Thus, if $(\Phi, k, a=0, b)$ is a "yes" instance, (\ref{eq:exp})
is equal to zero and we get to the right answer with probability
$1$. On the other hand, if it is a "no" instance then
$$\frac{1}{{2\choose k}}\sum_{i,j} tr\big(S\, \Phi^{\otimes 2
}(\rho^{i,j})\big)\geq \frac{1}{{2\choose k}}b,$$ and with
probability at least $\frac{1}{{2\choose k}}b$ which is greater
than an inverse polynomial, we get to a collision. Therefore,
$(\Phi, k, a=0, b)$ is in $\QMA_1$.

It remains to show that quantum clique, in the special case
stated in the theorem, is $\QMA_1$-hard. By theorem \ref{sat},
quantum $4$-SAT is $\QMA_1$-complete. Thus, if we establish a
polynomial time reduction from quantum $4$-SAT to quantum clique,
we are done.

Let $\big(\Pi_1, \dots \Pi_s, \epsilon \big)$ be an instance of
quantum $4$-SAT problem. Define the channel $\Phi$ as follows

$$\Phi(\rho)= \frac{1}{s}tr\big(\Pi\otimes I\, \rho\big)\vert 00\rangle\langle 00\vert
              +tr\big(M\otimes \vert 0\rangle\langle 0\vert\,\rho\big) \vert 11\rangle\langle 11\vert
              +tr\big(M\otimes \vert 1\rangle\langle 1\vert\,\rho\big) \vert 10\rangle\langle 10\vert,$$
where $\Pi=\sum_i \Pi_i$, and $M=I-\frac{1}{s}\Pi$, and consider
the instance $(\Phi, k=2, a=0, \frac{1}{s^2}\epsilon)$ of the
quantum clique problem. Note that, $\Phi$ is a q-c channel and by
lemma  \ref{lem:exact} it can be implemented exactly by a quantum
verifier. So $(\Phi, k=2, a=0, \frac{1}{s^2}\epsilon)$ satisfies
the conditions of theorem. The other parts of proof, that
$\big(\Pi_1, \dots \Pi_s, \epsilon \big)$ is a "yes" instance if
and only $(\Phi, k=2, a=0, \frac{1}{s^2}\epsilon)$, are exactly
same as in the proof of theorem \ref{thm:qclique}.

\end{proof}

\section{Complexity of computing Holevo capacity}

Here is the main theorem of this section.

\begin{thm} \label{thm:hcapacity} Suppose $\Phi$ is a quantum channel that
acts on an $n$-dimensional Hilbert space, and is given by
$poly(n)$ number of bits. Also, let $c$ be a real number. Then
deciding whether $\chi(\Phi)> c$, is $\NP$-complete.

\end{thm}

To prove this theorem we show that this problem is "harder" that
the problem of computing the minimum entropy of quantum channels,
and then prove computing the minimum entropy is $\NP$-complete.
In fact, the minimum entropy of channel, equation (\ref{eq:min}),
seems to be more tractable than the Holevo capacity. Then proving
the $\NP$-completeness of this problem is simpler.

\begin{thm}\label{thm:minent} Assume $\Phi$ is a quantum channel
acting on an $n$-dimensional Hilbert space, and is given by
polynomially many bits. Also, let $c$ be a real number. Then
deciding whether the minimum entropy of $\Phi$ is less than $c$
is $\NP$-complete.

\end{thm}

First, using theorem \ref{thm:minent} we prove theorem
\ref{thm:hcapacity}, and then get to the $\NP$-completeness of
computing minimum entropy.

\vspace{.6cm}

\noindent {\bf Proof of theorem \ref{thm:hcapacity}:} First of
all if $\Phi$ is a channel and $\chi(\Phi)\geq c$, then there are
probability distribution $\{p_i\}$ and states $\rho_1, \dots
\rho_s$ such that \begin{equation}\label{eq:check}H(\sum_i
p_i\Phi(\rho_i))-\sum_i p_iH(\Phi(\rho_i))\geq c. \end{equation}
The point is that, we may assume $s\leq n^2$, see \cite{chuang}.
Therefore, given this probability distribution and the quantum
states, the verifier can check whether (\ref{eq:check}) holds or
not. So, it is a problem in $\NP$.

To prove the hardness, since, by theorem \ref{thm:minent}
computing the minimum entropy is $\NP$-complete, if we establish a
reduction from minimum entropy to computing Holevo capacity we
are done.

Let $(\Phi, c)$ be an instance of minimum entropy problem as in
theorem \ref{thm:minent}. Let $\vert 1\rangle, \dots \vert
n\rangle$ be an orthonormal basis for the Hilbert space. Also, let
$X_0, \dots X_{n^2-1}$ be the $n$-dimensional generalized Pauli
matrices. That is, $X_{mn+d}=T^mR^d$, where $T\vert j\rangle=
\vert j+1 \mod n\rangle$ and $R\vert j\rangle=e^{2ij\pi/n}\vert
j\rangle$. Define the channel $\Psi$ such that

$$\Psi (\rho\otimes \vert i\rangle\langle i\vert )=X_i
\Phi(\rho)X_i^\dagger.$$ It is obvious that \begin{equation}
\label{eq:minmax}\chi(\Psi)=\max_{p_i, \rho_i} H\big( \sum_i\,
p_i\Psi(\rho_i) \big) -\sum_i\, p_iH(\Psi(\rho_i))\leq \log
n-\min_\rho H(\Psi(\rho)). \end{equation} Also, it is easy to see
that the minimum entropy of $\Psi$ is equal to the minimum
entropy of $\Phi$. On the other hand, if the minimum entropy of
$\Phi$ is taken on $\vert \phi\rangle$, and we let $\rho_i=\vert
\phi\rangle\langle \phi\vert \otimes \vert i\rangle\langle
i\vert$ and $p_i=1/n$, for $i=1,\dots n$, then equality holds in
(\ref{eq:minmax}). It means that, the minimum entropy of $\Phi$
is less than $c$ if and only if the Holevo capacity of $\Psi$ is
greater than $\log n-c$. We are done

\hfill$\Box$


\subsection{Complexity of computing the minimum capacity of a quantum channel}

The only remaining step is the proof of theorem \ref{thm:minent}.
To get a clearer proof it would be helpful to first state some
lemmas.

\subsubsection{Some lemmas on the minimum entropy of channels}

In this section we study some properties of the points that a
channel achieves its minimum entropy. Before stating the lemmas,
remember that the von Neumann entropy is convex, and then the
minimum entropy of a channel is attained on pure states.

\begin{lem}\label{lem:sum} Suppose $\Phi_1,\dots \Phi_k$ are $k$
channels with the same input and output state spaces. Also, assume
that the output states of every two of them are orthogonal. In
other words, for any $i,j$, $1\leq i<j\leq k$, and any states
$\rho$, $\rho'$,
$$tr(\Phi_i(\rho)\Phi_j(\rho'))=0.$$
Then
$$\min_{\rho} H\big(\sum_{i=1}^{k}\, p_i\Phi_i(\rho)\big) \geq
\sum_{i=1}^k p_i\min_\rho H(\Phi_i(\rho))+ H(p_1,\dots p_k),
$$
where, $\{p_1,\dots p_k\}$ is a probability distribution and
$H(p_1,\dots p_k)$ is its entropy. In particular, the minimum
entropy of $\sum_{i=1}^{k}\, p_i\Phi_i$ is at least $H(p_1, \dots
p_k)$, and equality holds iff there is $\rho$ such that all
states $\Phi_i(\rho)$ are pure.

\end{lem}

\begin{proof}
Since $\Phi_i(\rho)$'s have orthogonal supports
\begin{eqnarray*}
\min_{\rho} H\big(\sum_{i=1}^{k}\, p_i\Phi_i(\rho)\big) & = &
\min_{\rho} \sum_i -tr\big( p_i\Phi_i(\rho)\log(p_i\Phi_i(\rho))
\big)\\
                                                        & = &
\min_{\rho} \sum_i -p_i\, tr\big(\Phi_i(\rho) \log\Phi_i(\rho) +
\log p_i \Phi_i(\rho) \big) \\
                                                        & = &
\min_{\rho} \sum_i p_i H(\Phi_i(\rho)) -\sum_i p_i\log p_i \\
                                                        & \geq &
\sum_{i=1}^k p_i\min_\rho H(\Phi_i(\rho))+ H(p_1,\dots p_k).\\
\end{eqnarray*}

\end{proof}

\begin{lem} \label{lem:trace} Let $\Phi_{trace}$ be the channel that acts on
the Hilbert space $\mathcal{H}\otimes \mathcal{H}$, and traces
out the second register:
\begin{equation}\label{eq:trace}
\Phi_{trace}(\rho^{12})=tr_2(\rho^{12})=\rho^1.
\end{equation}
Then, the minimum entropy of $\Phi_{trace}$ is zero, and it is
achieved at the product states $\vert \psi_1\rangle\vert
\psi_2\rangle$.
\end{lem}

\begin{proof} Let $\vert \psi_{12}\rangle$ be a pure state in $\mathcal{H}\otimes \mathcal{H}$.
By the Schmidt decomposition \cite{chuang}, there are orthonormal
bases $\{\vert i\rangle\}$, $\{\vert i'\rangle \}$, and real
non-negative numbers $\lambda_i$, such that
\begin{equation}\label{eq:schmidt} \vert \psi_{12}\rangle=\sum_i \lambda_i \vert i\rangle\vert
i'\rangle.
\end{equation}
Hence, $\Phi(\vert \psi_{12}\rangle)=\sum_i \lambda_i^2\vert
i\rangle\langle i\vert$, and it is a pure state if only if just
one of $\lambda_i$'s is non-zero, or equivalently $\vert
\psi_{12}\rangle$ is a product state.

\end{proof}

The next lemma is on the minimum entropy of the SWAP test,
described in section \ref{subsec:swap}.

\begin{lem} \label{lem:swap} Let $\Phi_{swap}$ be the channel defined in
equation (\ref{eq:swaptest}). Then the minimum entropy of the
channel
$$\Phi(\rho)=\frac{1}{2}\Phi_{trace}(\rho)\otimes \vert u\rangle\langle u\vert\otimes \vert
000\rangle\langle 000\vert+\frac{1}{2}\vert u'_{12}\rangle\langle
u'_{12}\vert \otimes \vert 10\rangle\langle 10\vert\otimes
\Phi_{swap}(\rho),$$ where $\vert u \rangle\in \mathcal{H}$ and
$\vert u'_{12}\rangle\in \mathcal{H}\otimes \mathcal{H}$ are
arbitrary states, is equal to $H(2)=1$, and is attained at the
pure states of the form $\vert \psi\rangle \vert \psi\rangle$.
\end{lem}

\begin{proof} By lemma \ref{lem:sum}, it is sufficient to show
that
states of form $\vert \psi\rangle \vert \psi\rangle$ are the only
states $\rho$ such that $\Phi_{trace}(\rho)$ and
$\Phi_{swap}(\rho)$ are simultaneously pure.

Using lemma \ref{lem:trace}, such a state $\rho$ should be a
product state $\vert \psi_1\rangle \vert \psi_2\rangle$. On the
other hand, by equation (\ref{eq:swaptest}), it is clear that
$\Phi_{swap}(\vert\psi_1\rangle\vert\psi_2\rangle)$ is pure iff
$\vert \langle\psi_1\vert \psi_2\rangle\vert =1$, or equivalently
$\vert \psi_1\rangle=\vert \psi_2\rangle$.

\end{proof}

For the next lemma, it is helpful to fix some notations. Let
$\mathcal{H}$ an $n$-dimensional Hilbert space with the
orthonormal basis $\{\,\vert 1\rangle, \dots \vert n\rangle \}$.
For any $1\leq i< j\leq n$, let $\Pi_{ij}$ be the projection over
$\frac{1}{\sqrt{2}}(\vert i\rangle+\vert j\rangle)$,
$$\Pi_{ij}=\frac{1}{2}(\vert i\rangle+\vert j\rangle)(\langle
i\vert+\langle j\vert).$$ Also, let $\Pi'_{ij}$ be the projection
over $\frac{1}{\sqrt{2}}(\vert i\rangle-\vert j\rangle)$,
$$\Pi'_{ij}=\frac{1}{2}(\vert i\rangle-\vert j\rangle)(\langle
i\vert-\langle j\vert).$$

$\Pi_{ij}\otimes \Pi'_{ij}$ is a projection and its operator norm
is equal to one. Therefore, $\sum_{ij} \Pi_{ij}\otimes \Pi'_{ij}$
is a hermitian matrix and its norm is at most ${n\choose 2}$.
Thus,

$$M=I\otimes I-\frac{1}{n(n-1)}\sum_{ij} \Pi_{ij}\otimes \Pi'_{ij},$$
is a positive semidefinite matrix, and does not have zero
eigenvalue. It means that, $\vert v'_{12}\rangle$ is always in the
support of the following channel.

\begin{equation} \label{eq:cube}
\Phi_{cube}(\rho)= \frac{1}{n(n-1)} \sum_{ij}
tr\big(\Pi_{ij}\otimes \Pi'_{ij}\, \rho \big)\vert
v_{12}\rangle\langle v_{12}\vert + tr\big( M\rho \big)\vert
v'_{12}\rangle\langle v'_{12}\vert.
\end{equation}

\begin{lem}\label{lem:cube} Let $\vert v_{12}\rangle,\vert v'_{12}\rangle\in
\mathcal{H}\otimes \mathcal{H}$ be two orthogonal states, and
define $\Phi_{cube}$ as in equation (\ref{eq:cube}). Then the
minimum entropy of the channel

\begin{eqnarray*} \Phi(\rho)& = & \frac{1}{3}\Phi_{trace}(\rho)\otimes \vert
u\rangle\langle u\vert\otimes \vert 000\rangle\langle
000\vert+\frac{1}{3}\vert u'_{12}\rangle\langle u'_{12}\vert
\otimes \vert 10\rangle\langle 10\vert\otimes \Phi_{swap}(\rho)
\\ & & + \frac{1}{3} \Phi_{cube}(\rho)\otimes \vert 110\rangle\langle
110\vert \end{eqnarray*} is equal to $H(3)=\log 3$ and is attained
at the states $\vert \psi_{12}\rangle =\vert \psi\rangle\vert
\psi\rangle$, where
\begin{equation}\label{eq:states}
\vert \psi\rangle = \frac{1}{\sqrt{n}}\sum_{i=1}^n\, x_i\vert
i\rangle ,
\end{equation}
and $x_i\in \{+1, -1\}$.

\end{lem}

\begin{proof} Again, using lemma \ref{lem:sum}, it suffices to
show that the only states $\rho$ such that $\Phi_{trace}(\rho)$,
$\Phi_{swap}(\rho)$ and $\Phi_{cube}(\rho)$ are pure, are the
states $\vert \psi\rangle \vert \psi\rangle$ where
$\vert\psi\rangle$ is of the form (\ref{eq:states}).

In lemma \ref{lem:swap}, we showed if $\Phi_{trace}(\rho)$ and
$\Phi_{swap}(\rho)$ are pure then $\rho$ is a pure state of the
form $\vert\psi\rangle\vert\psi\rangle$. So, it remains to show
that if $\Phi_{cube}(\vert\psi\rangle\vert\psi\rangle)$ is pure
then $\vert\psi\rangle$ is of the form (\ref{eq:states}).

As we mentioned, $\vert v'_{12}\rangle$ is always in the support
of $\Phi_{cube}(\rho)$. Hence, if
$\Phi_{cube}(\vert\psi\rangle\vert\psi\rangle)$ is pure then it
is equal to $\vert v'_{12}\rangle$. It means that,
$\Phi_{cube}(\rho)$ is pure if and only if $$tr\big(
\Pi_{ij}\otimes \Pi'_{ij}\, \rho \big)=0,$$ for any $i,j$. Let
$\vert\psi\rangle= \sum_i \lambda_i\vert i\rangle$, and suppose
$\Phi_{cube}(\vert\psi\rangle\vert\psi\rangle)$ is pure. We have
$$0=\langle \psi\vert\langle\psi\vert \Pi_{ij}\otimes \Pi'_{ij}\vert\psi\rangle\vert\psi\rangle
= \langle \psi\vert \Pi_{ij}\vert \psi\rangle\langle \psi\vert
\Pi'_{ij}\vert \psi\rangle=\frac{1}{4}\vert
\lambda_i+\lambda_j\vert^2 \vert \lambda_i-\lambda_j\vert^2.$$ In
other words, for any $i,j$, either $\lambda_i=\lambda_j$ or
$\lambda_i=-\lambda_j$. So, $\vert \psi\rangle$ should be of the
form (\ref{eq:states}).

\end{proof}

\subsubsection{Proof of theorem \ref{thm:minent}}

To prove the $\NP$-hardness of the problem of computing minimum
entropy, we should find a reduction from an $\NP$-complete
problem to this one. The most comfortable such problem for us is
the 2-Out-of-4-SAT problem \cite{khanna}. We can formulate this
problem as follows. Given $m=poly(n)$ vectors of the form

$$\vert A_k\rangle =\sum_{i=1}^n a^k_i\, \vert i\rangle,$$
where for each $k$, $1\leq k\leq m$, there are exactly four
non-zero $a^k_i$, and $a^k_i$ is zero or $\pm \frac{1}{2}$,
decide whether there exists a vector $\vert \psi\rangle$ of the
form (\ref{eq:states}) orthogonal to all $\vert A_k\rangle$'s,
$\langle A_k\vert \psi\rangle=0$.

Now we are ready to prove the theorem. Given a witness state
$\rho$, we can check whether $H(\Phi(\rho))< c$, in polynomial
time. Therefore, this problem is in $\NP$.

To prove hardness, let $\vert A_1\rangle,\dots \vert A_m\rangle$
be an instance of 2-Out-of-4-SAT. Let

$$H=\frac{1}{m}\sum_{k=1}^m \vert A_k \rangle\langle A_k\vert\otimes \vert A_k \rangle\langle A_k\vert$$
and define
$$\Phi_H(\rho)= \frac{1}{2}tr(H\rho) \vert w_{12}\rangle
\langle w_{12}\vert + tr\big((I\otimes I-\frac{1}{2}H)\,
\rho\big)\vert w'_{12}\rangle \langle w'_{12}\vert,
$$
where $\vert w_{12}\rangle$ and $\vert w'_{12}\rangle$ are two
orthogonal states in $\mathcal{H}\otimes \mathcal{H}$. Since the
norm of $\frac{1}{2}H$ is less than or equal to $1/2$, $\vert
w'_{12}\rangle$ is is always in the support of $\Phi_H$. Then the
minimum entropy of $\Phi_H$ is zero and is achieved at the states
$\vert\psi_{12}\rangle$ that are orthogonal to all $\vert
A_k\rangle \vert A_k\rangle$, $k=1,\dots ,m$.

Now, define the channel

\begin{eqnarray*} \label{eq:mainchannel} \Phi(\rho) & = & \frac{1}{4}\Phi_{trace}(\rho)\otimes \vert
u\rangle\langle u\vert\otimes \vert 000\rangle\langle
000\vert+\frac{1}{4}\vert u'_{12}\rangle\langle u'_{12}\vert
\otimes \vert 10\rangle\langle 10\vert\otimes \Phi_{swap}(\rho)
\\ & & + \frac{1}{4} \Phi_{cube}(\rho)\otimes \vert 110\rangle\langle
110\vert+ \frac{1}{4}\Phi_{H}(\rho)\otimes \vert 111\rangle
\langle 111\vert.
\end{eqnarray*}
By lemma \ref{lem:sum}, the minimum entropy of $\Phi$ is at least
$H(4)=2$, and equality holds if there exists $\vert
\psi_{12}\rangle$ such that all the states $\Phi_{trace}(\vert
\psi_{12})$, $\Phi_{swap}(\vert \psi_{12})$, $\Phi_{cube}(\vert
\psi_{12})$ and $\Phi_{H}(\vert \psi_{12})$ are pure. By lemma
\ref{lem:cube} such a state should be of the form $\vert
\psi_{12}\rangle=\vert \psi\rangle\vert \psi\rangle$, where
$\vert \psi\rangle$ is of the form (\ref{eq:states}). Also, for
such a state $\Phi_H(\vert \psi\rangle\vert \psi\rangle)$ is pure
iff $\langle \psi\vert A_k\rangle=0$, $k=1,\dots, m$. Therefore,
the minimum entropy of $\Phi$ is $H(4)$, if and only if $(\vert
A_1\rangle, \dots \vert A_m\rangle)$ is a "yes" instance of
2-Out-of-4-SAT problem. Also, using the integrality of the
problem there exists $\epsilon
> 1/poly(n)$, such that $(\vert
A_1\rangle, \dots \vert A_m\rangle)$ is a "yes" instance if and
only if the minimum entropy of $\Phi$ is less than $2+\epsilon$.
Thus, computing the minimum entropy is $\NP$-complete.

\subsection{Computing Holevo capacity of entanglement breaking channels}

We proved that computing the minimum entropy, and then, Holevo
capacity are $\NP$-complete. In these two theorems, we considered
general quantum channels. But one suspects that if we restrict
ourselves to a special class of quantum channels then we get to
simpler problems.

For example, let $\Phi$ be a classical-quantum channel (c-q
channel)
\begin{equation} \label{eq:c-q} \Phi(\rho)=\sum_{i=1}^n \langle
i\vert \rho\vert i\rangle \sigma_i ,
\end{equation}
where $\vert 1\rangle,\dots \vert n\rangle$ is an orthonormal
basis and $\sigma_1,\dots \sigma_n$ are arbitrary states. Then,
obviously, the minimum entropy of $\Phi$ is equal to $$\min_i
H(\sigma_i),$$ and then, can be computed in polynomial time. Also,
it is easy to see that computing the Holevo capacity of $\Phi$ is
a convex optimization problem and can be solved efficiently.

Therefore, to get a non-trivial problem we should consider a more
general class of quantum channels. Indeed, c-q channels that we
considered in equation (\ref{eq:c-q}), and also q-c channels,
equation (\ref{eq:q-c}), are special cases of {\it Entanglement
breaking channels}. An entanglement breaking channel is a channel
$\Phi$ of the form
\begin{equation}\label{eq:entbreak} \Phi(\rho)=\sum_{i=1}^r
tr(M_i\rho)\sigma_i,
\end{equation}
where $\{M_i\}$ is a POVM and $\sigma_1,\dots \sigma_r$ are
arbitrary states. Although, it seems that the problem of computing
the minimum entropy and Holevo capacity of entanglement breaking
channels is simpler than the general case, we prove that these are
also $\NP$-complete.

\begin{thm}\label{thm:entbreak} Assume $\Phi$ is an entanglement
breaking channel of the form (\ref{eq:entbreak}) acting on an
$n$-dimensional Hilbert space, and is given by polynomially many
bits. Also let $c$ be a real number. Then the questions of
bounding the Holevo capacity and the minimum entropy of $\Phi$,
$$\chi(\Phi)> c$$
and
$$\min_{\rho} H(\Phi(\rho))< c,$$
are $\NP$-complete.

\end{thm}

Note that, if we show that computing the minimum entropy for
entanglement breaking channels is $\NP$-hard, then by the same
argument as in the proof of theorem \ref{thm:hcapacity} we can
prove the hardness of computing the Holevo capacity. Also, recall
that, in the proof of theorem \ref{thm:minent} all the channels
that we used were entanglement breaking except $\Phi_{trace}$.
Therefore, if we replace $\Phi_{trace}$ with an entanglement
breaking channel that captures the same properties then we are
done.

The key idea is the following observation first appeared in
\cite{jozsa}. Suppose $\rho$ is the density matrix of a two-qubit
state. Let $\sigma_0=I, \sigma_1, \sigma_2, \sigma_4$ be the
Pauli matrices. Also, for $i=1,2,3$ let $P_i^\pm=\frac{1}{2}(I\pm
\sigma_i)$ be density matrices of the $+1$ and $-1$ eigenstates
of $\sigma_i$. For $0\leq i,j\leq 3$ define
$c_{ij}=tr(\sigma_i\otimes \sigma_j \rho)$. Then, we know that
$$\rho=\frac{1}{4}\sum_{i,j=0}^3 c_{ij}\, \sigma_i\otimes
\sigma_j.$$ If we rewrite this equation in terms of $P_i^{\pm}$,
we get to
\begin{eqnarray*} \rho  =  \frac{1}{4}\sum_{i,j=1}^3  &
(\frac{1}{9}+\frac{1}{3}c_{i0}+\frac{1}{3}c_{0j}+c_{ij})P_i^+\otimes
P_j^+ \\
 & +
(\frac{1}{9}-\frac{1}{3}c_{i0}+\frac{1}{3}c_{0j}-c_{ij})P_i^-\otimes
P_j^+\\
 &  +
(\frac{1}{9}+\frac{1}{3}c_{i0}-\frac{1}{3}c_{0j}-c_{ij})P_i^+\otimes
P_j^-\\
 &  +
(\frac{1}{9}-\frac{1}{3}c_{i0}-\frac{1}{3}c_{0j}+c_{ij})P_i^-\otimes
P_j^-.
\end{eqnarray*}
Suppose all the coefficients in this expression are non-negative.
Then
\begin{eqnarray*} tr_2(\rho)  =  \frac{1}{4}\sum_{i,j=1}^3  &
(\frac{1}{9}+\frac{1}{3}c_{i0}+\frac{1}{3}c_{0j}+c_{ij})P_i^+ \\
 & +
(\frac{1}{9}-\frac{1}{3}c_{i0}+\frac{1}{3}c_{0j}-c_{ij})P_i^-\\
 &  +
(\frac{1}{9}+\frac{1}{3}c_{i0}-\frac{1}{3}c_{0j}-c_{ij})P_i^+\\
 &  +
(\frac{1}{9}-\frac{1}{3}c_{i0}-\frac{1}{3}c_{0j}+c_{ij})P_i^-.\\
\end{eqnarray*}
In other words, $tr_2(\rho)$ can be written as a linear
combination of states $P_i^{\pm}$ with coefficients of the form
$tr(M\rho)$, where $\{M\}$ is some POVM.

It means that, if the coefficients were always non-negative then
$\rho\mapsto tr_2(\rho)$ was an entanglement breaking channel. To
satisfy this extra assumption we can replace $\rho$ with
$\rho_\epsilon= 1/4(1-\epsilon)I+\epsilon \rho $, where
$0<\epsilon<1/16 $, and observe that the coefficients for
$\rho_{\epsilon}$ are all non-negative. In general, we have the
following lemma, proved in \cite{jozsa}.

\begin{lem}\label{lem:un-ent} Let $\rho$ be a state in $\mathcal{H}\otimes
\mathcal{H}$ where $\mathcal{H}$ is an $n$-dimensional Hilbert
space. Also let $1/n^2I$ be the maximally mixed state in
$\mathcal{H}\otimes \mathcal{H}$ and $0<\epsilon <1/n^2$. Then
$1/n^2(1-\epsilon)I+\epsilon\rho$ is an un-entangled state. Also,
as a consequence,
\begin{equation}\label{eq:trace'}\Phi'_{trace}(\rho)=tr_2(1/n^2(1-\epsilon)I+\epsilon\rho)\end{equation}
is an entanglement breaking channel.
\end{lem}

Using this lemma, the proof of theorem \ref{thm:entbreak} follows
immediately.

\vspace{.6cm}

\noindent{\bf Proof of theorem \ref{thm:entbreak}:} All the steps
of the proof are same as in theorem \ref{thm:minent}, except that
we replace the channel $\Phi_{trace}$ with $\Phi'_{trace}$, which
is an entanglement breaking one. The only property we should check
is that the minimum entropy of $\Phi'_{trace}$ is achieved at
product states. It holds because
$tr_2(1/n^2(1-\epsilon)I+\epsilon \rho)=1/n(1-\epsilon)I+\epsilon
tr_2(\rho)$, and $H(tr_2(1/n^2(1-\epsilon)I+\epsilon \rho))$ is
minimum if and only if $H(tr_2(\rho))$ is minimum.

\hfill$\Box$

\section{Conclusion}

In this paper we prove that the quantum clique problem is
$\QMA$-complete. This is obtained by considering an $\NP$-complete
problem, and somehow translating it to the language of quantum
information theory. The key point is that clique problem in graphs
can be stated in terms of zero error capacity. So, this
translation is straightforward. Now the question is that whether
this method can lead us to other $\QMA$-complete problems. Note
that this idea is first captured in the $\QMA$-completeness of
local Hamiltonian problem by considering $\NP$-harness of SAT.

In the second part of paper, we consider the problem of computing
the Holevo capacity, and then, minimum entropy of a quantum
channel, and prove that they are $\NP$-complete. Since, there are
few results on the computational complexity of invariants of
quantum channel, it would be a natural question to consider the
complexity of other such quantities for channels as well as
quantum states.

\end{document}